\documentclass{article}
\usepackage{spconf,amsmath,graphicx}

\usepackage{enumitem}
\setlist{nosep, leftmargin=14pt}
\usepackage{subfigure}
\usepackage{mwe} % to get dummy images
\usepackage[T1]{fontenc}
\usepackage{multirow}
\usepackage{amsmath,amsfonts,amssymb}
\usepackage{makecell}
\usepackage[colorlinks=true, allcolors=blue]{hyperref}
\usepackage{diagbox}
\title{Metal Inpainting in CBCT Projections Using Score-based Generative Model}
\name{Siyuan Mei, Fuxin Fan, Andreas Maier}
\address{Friedrich-Alexander-Universit\"at Erlangen-N\"urnberg\\
Pattern Recognition Lab\\
Erlangen, Germany}

\begin{document}
%\ninept
%
\maketitle
\begin{abstract}
During orthopaedic surgery, the inserting of metallic implants or screws are often performed under mobile C-arm systems. Due to the high attenuation of metals, severe metal artifacts occur in 3D reconstructions, which degrade the image quality greatly. To reduce the artifacts, many metal artifact reduction algorithms have been developed and metal inpainting in projection domain is an essential step. In this work, a score-based generative model is trained on simulated knee projections and the inpainted image is obtained by removing the noise in conditional resampling process. The result implies that the inpainted images by score-based generative model have more detailed information and achieve the lowest mean absolute error and the highest peak-signal-to-noise-ratio compared with interpolation and CNN based method. Besides, the score-based model can also recover projections with big circlar and rectangular masks, showing its generalization in inpainting task.
\end{abstract}
\section{Introduction}
\label{sec:intro}

The mobile C-arm systems are frequently used in interventional surgery currently, which supports the accurate placement of metallic impalnts or screws. However, some physical effects like photon starvation and beam hardening occur when x-rays pass through metals, resulting the bright and dark streak artifacts, which reduce the quality of the reconstructed image~\cite{katsura2018current}. The conventional metal artifacts reduction (MAR) algorithms tackle the problem by sinogram completion~\cite{meyer2010normalized,meyer2012frequency} and iterative reconstruction~\cite{wang1996iterative,zhang2011metal}. Deep learning based methods are also applied in MAR~\cite{zhang2018convolutional,liao2019generative,gottschalk2019deep,wang2021dual,wang2021dicdnet} and the models are trained on paired data in the supervised way. For the C-arm system, since only the central slice in cone-beam geometry can be represented as sinogram and other artifacts like truncation artifacts also exist in reconstructions, image inpainting in projection domain is the feasible way for MAR.

Score-based generative models are applied in computer vision recently~\cite{song2020score,ho2020denoising} and with the mechanism of stepwise noise perturbation and removal, they have shown superiority over generative adversarial networks~\cite{dhariwal2021diffusion}. Such models are successfully applied in image inpainting task~\cite{song2020score,lugmayr2022repaint}. In Ref.~\cite{lugmayr2022repaint}, a score-based generative model called RePaint is trained and it can generate restored image with high fidelity under different masks. Besides, score-based generative models are used in the field of medical imaging processing, such as CT and MRI reconstruction~\cite{song2021solving,chung2022score}. Inspired by the research above, a score-based model is trained on knee projections and this is the first study to apply such model in metal inpainting in CBCT projections. 

\section{Materials and method}
\label{sec:format}

\subsection{Score-based generative models}
\begin{figure*}[htb]
   \begin{center}
   \begin{tabular}{c} %% tabular useful for creating an array of images 
   \includegraphics[width=15cm]{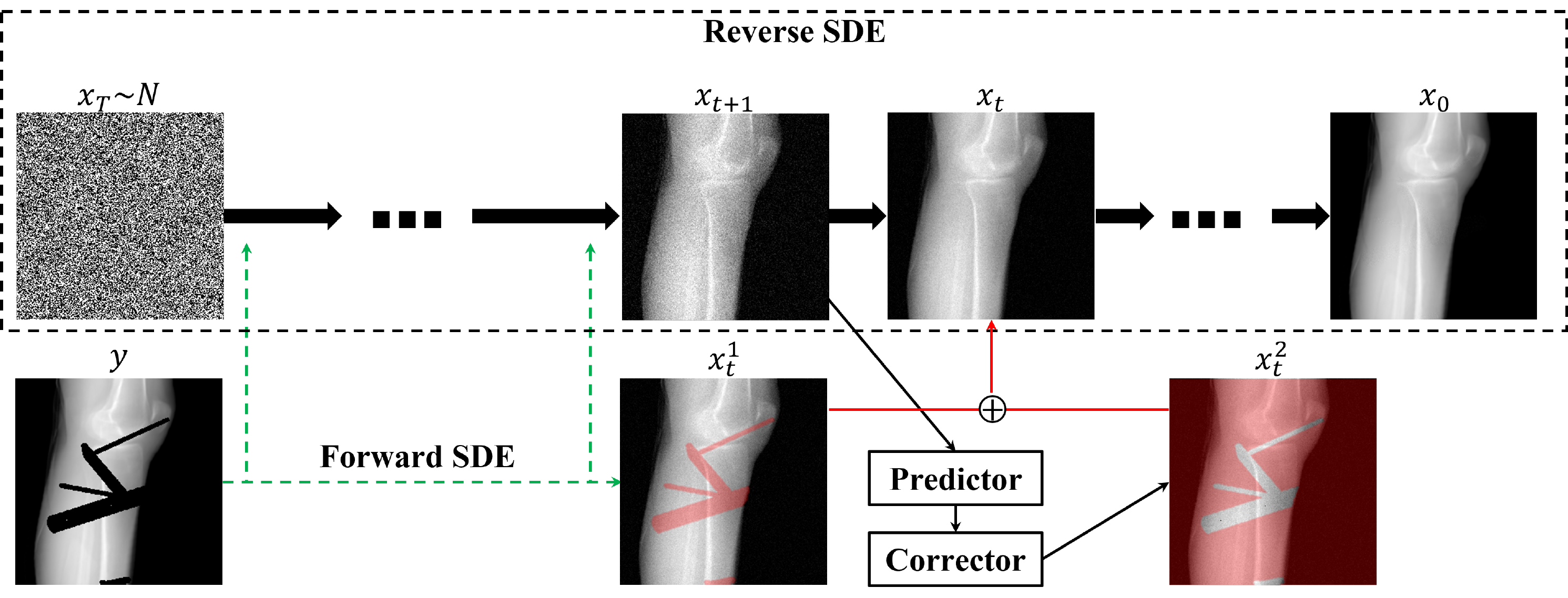}
   \end{tabular}
   \end{center}
   \caption[figure1] 
%>>>> use \label inside caption to get Fig. number with \ref{}
   { \label{fig:figure1} 
The illustration of the resampling process for projection inpainting.}
\end{figure*} 
Score-based generative models through stochastic differential equation (SDE) consist of three steps: perturbation process, reverse process and resampling. In the perturbation process, the original image $\boldsymbol{x}_0$ is perturbed with a stochastic process to construct a continuous diffusion process $\boldsymbol{x}_t$, where $t\in[0,1]$. The perturbation process by SDE has the following form
\begin{equation}
d\boldsymbol{x}_t=\boldsymbol{f}(\boldsymbol{x}_t,t)dt+g(t)d\boldsymbol{w}_t, \quad t\in[0,1]\, ,
\end{equation}
where $\boldsymbol{f}(\boldsymbol{x},t)$ is the drift coefficient, $g(t)$ is the diffusion coefficient and $\boldsymbol{w}_t$ denotes a standard Wiener process. Depending on the change of variance during the perturbation process, SDEs are clarified into variance exploding (VE), variance preserving (VP) and sub VP SDE. In this paper, VE SDE is used, which has the form of
\begin{equation}
\boldsymbol{f}=\boldsymbol{0}, g=\sqrt{\frac{d[\sigma^{2}(t)]}{dt}}\,,
\end{equation}
\begin{equation}
\sigma(t)=\sigma(0)(\frac{\sigma(1)}{\sigma(0)})^t, \quad t\in[0,1]\, ,
\end{equation}
where $\sigma(t)>0$ is increasing with the increment of $t$. Then the diffusion process $\boldsymbol{x}_t$ at time point $t$ can be obtained by  
\begin{equation}
\boldsymbol{x}_t=\boldsymbol{x}_0+[\sigma^{2}(t)-\sigma^{2}(0)]\textbf{I},
\end{equation}
where $\textbf{I}$ is a normal distribution. The reverse process follows the equation of reverse-time SDE
\begin{equation}
d\boldsymbol{x}_t=[\boldsymbol{f}(\boldsymbol{x}_t,t)dt-g(t)^{2}\triangledown_{\boldsymbol{x}_t}\log p_{t}(\boldsymbol{x}_t)]+g(t)d\boldsymbol{\bar{w}}_t, \\ t\in[0,1]\,,
\label{equ_reverse}
\end{equation}
where $\triangledown_{\boldsymbol{x}_t}\log p_{t}(\boldsymbol{x}_t)$ is the score function of $ p_{t}(\boldsymbol{x}_t)$ and $\boldsymbol{\bar{w}}_t$ denotes the reverse Wiener process. Following the reverse process, the perturbed image will be restored gradually. In the case of VE SDE which is used in this paper, the Eq.~\ref{equ_reverse} can be rewritten as
\begin{equation}
\label{equ_reverse}
d\boldsymbol{x}_t=-\frac{d[\sigma^{2}(t)]}{dt}\triangledown_{\boldsymbol{x}_t}\log p_{t}(\boldsymbol{x}_t)+\frac{d[\sigma^{2}(t)]}{dt}d\boldsymbol{\bar{w}}_t, \quad t\in[0,1]\,.
\end{equation}
Then $\boldsymbol{x}_t$ in the reverse process has the form of
\begin{equation}
\boldsymbol{x}_t=\boldsymbol{x}_{t+\Delta t}-d\boldsymbol{x}_{t+\Delta t}.
\end{equation}  
   
For projection inpainting, the missing pixels in the metal area should be restored and the pixels in the background serve as conditional information in the resampling process. The pipeline for projection inpainting is shown in Fig.~\ref{fig:figure1}. We denote the projection to be inpainted as $\boldsymbol{y}$ and the binary metal mask with ones in the background as $\boldsymbol{m}$. $\boldsymbol{x}_t^1$ is obtained by forward SDE and has the form of
\begin{equation}
\boldsymbol{x}_t^1=\boldsymbol{y}+[\sigma^{2}(t)-\sigma^{2}(0)]\textbf{I},
\end{equation}
and it provides the background information. $\boldsymbol{x}_t^2$ follows reverse SDE and is written as
\begin{equation}
\boldsymbol{x}_t^2=\boldsymbol{x}_{t+1}-d\boldsymbol{x}_{t+1},
\end{equation}
and it predicts the inpainted pixels. Then the restored projection $\boldsymbol{x}_t$ at time point $t$ can be obtained as
\begin{equation}
\boldsymbol{x}_t=\boldsymbol{x}_t^1\odot \boldsymbol{m} + \boldsymbol{x}_t^2\odot(1-\boldsymbol{m}),
\end{equation}
where $\odot$ means pixelwise multiplication.

\subsection{Data generation}
The knee CT volumes are selected from the whole body CT volumes from the SICAS medical image repository \cite{kistler2013virtual}. In total, there are 50 volumes with single leg for each of them and all volumes are rescaled to a voxel side length of 0.5\,mm. All the volumes are forward projected to generate projections by CONRAD \cite{maier2013conrad}. The parameters for defining the C-arm trajectory are listed in Tab.~\ref{tab:para}. 3000 projections are generated and 2700 of them are used for model training and 300 projections are used for test. To generate metal masks, some implants like K-wires, screws, plates with holes are drawn using the software AutoCAD. Then these implants are randomly selected and placed in different 3D volumes. The metal masks are obtained by forward projecting these multi-metal volumes under the same parameters shown in Tab.~\ref{tab:para}.

\begin{table}[htb]
\caption{Parameters for the mobile C-arm system.} 
\label{tab:para}
\begin{center}       
\begin{tabular}{|l|l|} %% this creates two columns
%% use of \rule[]{}{} below opens up each row
\hline
\rule[-1ex]{0pt}{3.5ex}  Parameter & Value  \\
\hline
\rule[-1ex]{0pt}{3.5ex}  Scan angular range & 360$^{\circ}$   \\
\hline
\rule[-1ex]{0pt}{3.5ex}  Incremental Angular step & 6$^{\circ}$   \\
\hline
\rule[-1ex]{0pt}{3.5ex}  Source-to-detector distance & 1164\,mm   \\
\hline
\rule[-1ex]{0pt}{3.5ex}  Source-to-isocentor distance & 622\,mm   \\
\hline
\rule[-1ex]{0pt}{3.5ex}  Detector Size & 256\,$\times$\,256   \\
\hline
\rule[-1ex]{0pt}{3.5ex}  Detector pixel size & 1.16 mm\,$\times$\,1.16 mm   \\
\hline
\end{tabular}
\end{center}
\end{table}

\subsection{Network Structures}

   \begin{figure}[htb]
   \begin{center}
   \begin{tabular}{c} %% tabular useful for creating an array of images 
   \includegraphics[width=8cm]{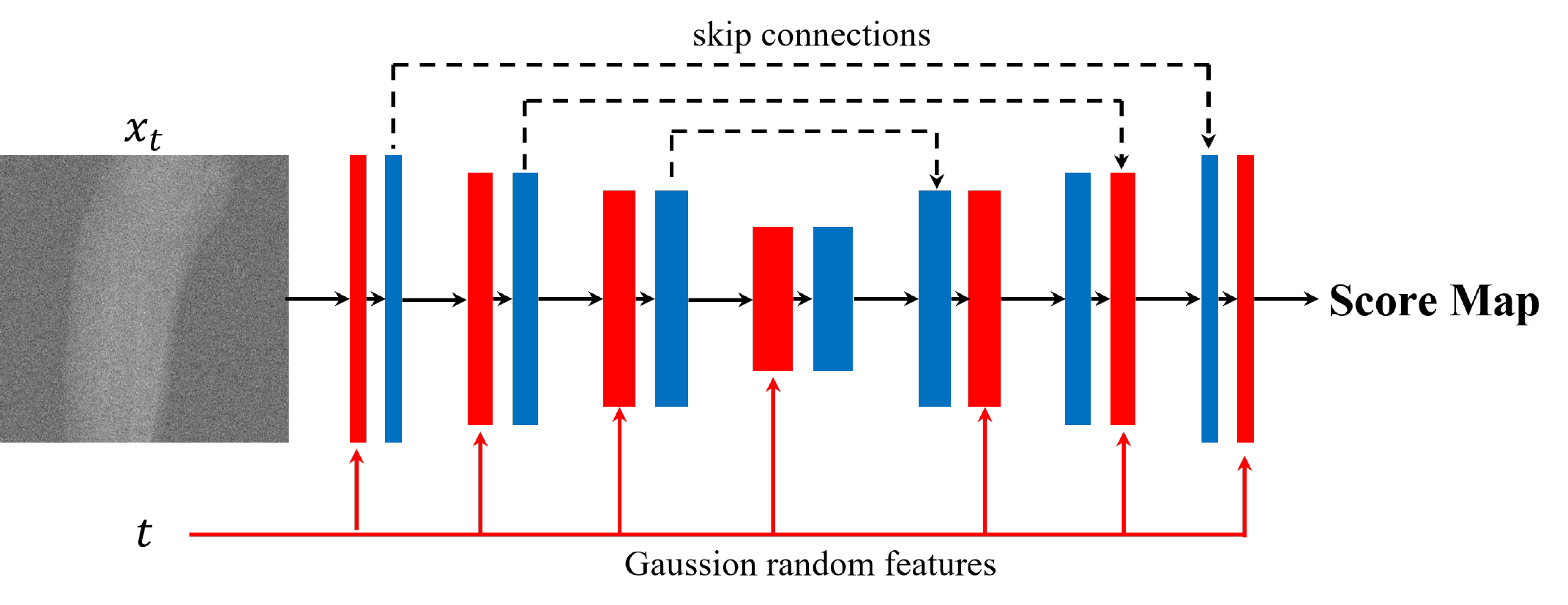}
   \end{tabular}
   \end{center}
   \caption[figure2] 
%>>>> use \label inside caption to get Fig. number with \ref{}
   { \label{fig:figure2} 
The network structure for the score-based generative model.}
   \end{figure}
   
Since the mask pyramid network (MPN) from Ref.~\cite{liao2019generative} has good performance in projection inpainting, its results are used for comparison. The projection to be inpainted and the corresponding metal mask are the inputs for MPN, both of which have the size of $256\,\times\,256$. According to Ref.~\cite{song2020score}, the network structure for the score-based generative model has high flexibility. As shown in Fig.~\ref{fig:figure2}, the structure of MPN without downsampling layers from metal mask serves as the backbone for the score-based generative model in this work. Besides, the Gaussian random features \cite{tancik2020fourier} are generated at time step $t$ and these features are summed up with the corresponding blocks from the network, which are labeled in red in Fig.~\ref{fig:figure2}. For the defined VE SDE in this work, the minimum and maximum variances are 0.01 and 128, respectively. During the resampling process, the image of gaussion noise with variance of $\sigma(1)$ is the initializer and the inpainted projection is generated by predictor-corrector (PC) resampler for 1000 steps.  All experiments are conducted on Nvidia A100 GPU.
% The hyperparameter optimization for the resamping process is included in~\ref{A}.

\subsection{Hyperparameter optimization}

\begin{table*}[htb!]
\caption{Ablation study on SNR and sampling steps under same PC sampler.}
\label{tab:hyper-parapeters}
\begin{center}   
\begin{tabular}{p{2.4cm}|p{1.2cm}|p{1.2cm}|p{1.2cm}|p{1.2cm}|p{1.2cm}|p{1.2cm}}
\hline
                  & \multicolumn{2}{c|}{$N$=500} & \multicolumn{2}{c|}{$N$=1000} & \multicolumn{2}{c}{$N$=2000} \\  \hline
Metric            & MAE         & PSNR        & MAE          & PSNR        & MAE  & PSNR  \\ \hline
% $\eta$=0.209         & 0.0197  & 40.83       & 0.0167& 42.05    & 0.0155 & 42.80 \\
$\eta$=0.20         & 0.0197  & 40.85       & 0.0172& 41.99    & 0.0154 & 42.74 \\
\hline
$\eta$=0.40*         & 0.0161 &  42.39       & 0.0148 & 43.00     & 0.0146 & 43.07 \\ \hline
$\eta$=0.60         & 0.0172 & 41.94       & 0.0178 & 41.67       & 0.0201 & 40.82 \\ \hline
time per slice(s) & \multicolumn{2}{c|}{2.50}  & \multicolumn{2}{c|}{4.38}   & \multicolumn{2}{c}{9.79} \\ \hline  
\end{tabular}
\end{center}
\end{table*}

Some hyperparameters in the sampling process are considered, including the signal-to-noise ratio (SNR) $\eta$ for corrector, and the number of discretization steps $N$ for reverse-time SDE. SNR determines the step size $\epsilon $ in Langevin dynamics, and $N$ corresponds to the noise scales. The quantitative results for 300 projections with metal masks under $ancestral$ $sampling$ predictor and $Langevin$ corrector are listed in Tab.~\ref{tab:hyper-parapeters}. It demonstrates that $\eta = 0.4$ (with star remarked) can achieve better performance than others. Time consumption is an important issue in clinical applications. According to Tab.~\ref{tab:hyper-parapeters}, $N$ is proportional to time usage, and it has limited performance improvement from 1000 to 2000 steps. Therefore, the parameters in our experiments are chosen as $\eta = 0.4$, $N$=1000.

\section{Results and discussion}
%   \begin{figure} [htb]
%   \begin{center}
%   \begin{tabular}{c} %% tabular useful for creating an array of images 
%   \includegraphics[width=7.5cm]{normalcase.pdf}
%   \end{tabular}
%   \end{center}
%   \caption[figure3] 
% %>>>> use \label inside caption to get Fig. number with \ref{}
%   { \label{fig:figure3} 
% The inpainting results under different metal masks.}
%   \end{figure} 
   
\begin{figure}[htb]
\centering

    \begin{minipage}[t]{0.18\linewidth}
    \centering
    \centerline{Label}\medskip
\end{minipage}
\hfill
    \begin{minipage}[t]{0.18\linewidth}
    \centering
    \centerline{Input}\medskip
\end{minipage}
\hfill
    \begin{minipage}[t]{0.18\linewidth}
    \centering
    \centerline{Interpolation}\medskip
\end{minipage}
\hfill
    \begin{minipage}[t]{0.18\linewidth}
    \centering
    \centerline{MPN}\medskip
\end{minipage}
\hfill
    \begin{minipage}[t]{0.18\linewidth}
    \centering
    \centerline{Ours}\medskip
\end{minipage}

    \begin{minipage}[b]{0.18\linewidth}
    \centering
    \centerline{\includegraphics[width=1.6cm]{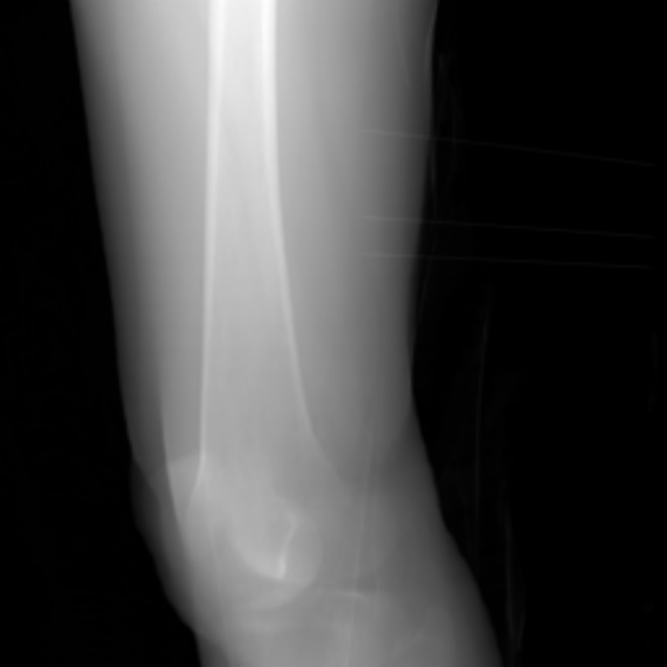}}
    \centerline{(a0)}\medskip
\end{minipage}
\hfill
    \begin{minipage}[b]{0.18\linewidth}
    \centering
    \centerline{\includegraphics[width=1.6cm]{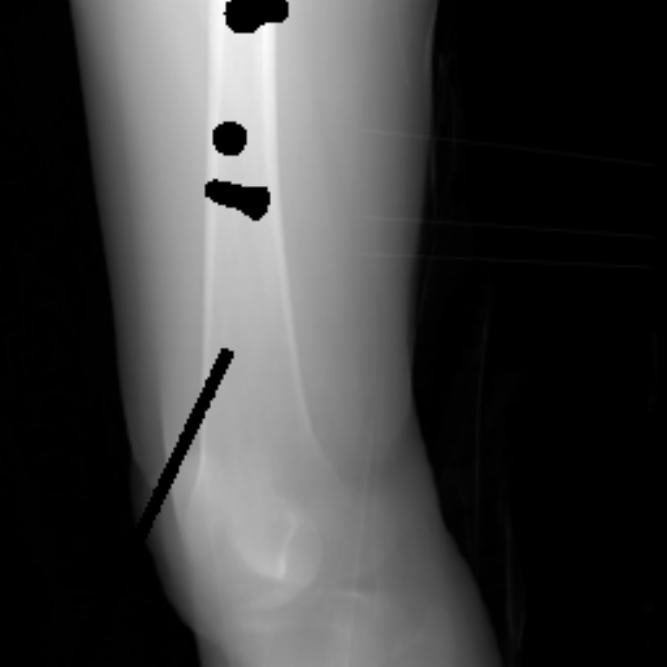}} 
    \centerline{(a1)}\medskip
\end{minipage}
\hfill
    \begin{minipage}[b]{0.18\linewidth}
    \centering
    \centerline{\includegraphics[width=1.6cm]{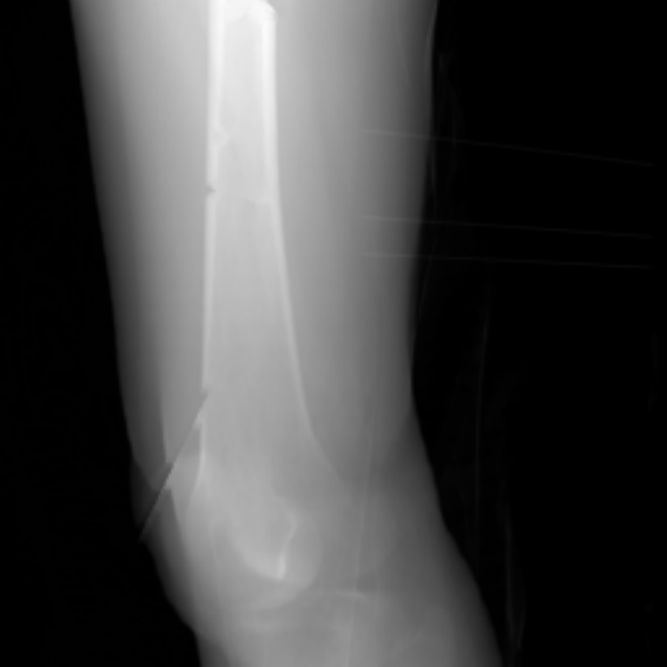}} 
    \centerline{(a2)}\medskip
\end{minipage}
\hfill
    \begin{minipage}[b]{0.18\linewidth}
    \centering
    \centerline{\includegraphics[width=1.6cm]{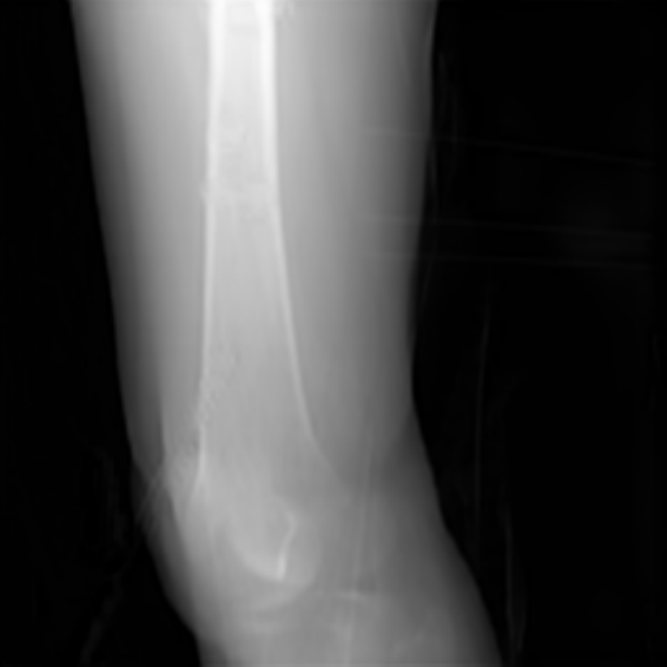}}
    \centerline{(a3)}\medskip
\end{minipage}
\hfill
    \begin{minipage}[b]{0.18\linewidth}
    \centering
    \centerline{\includegraphics[width=1.6cm]{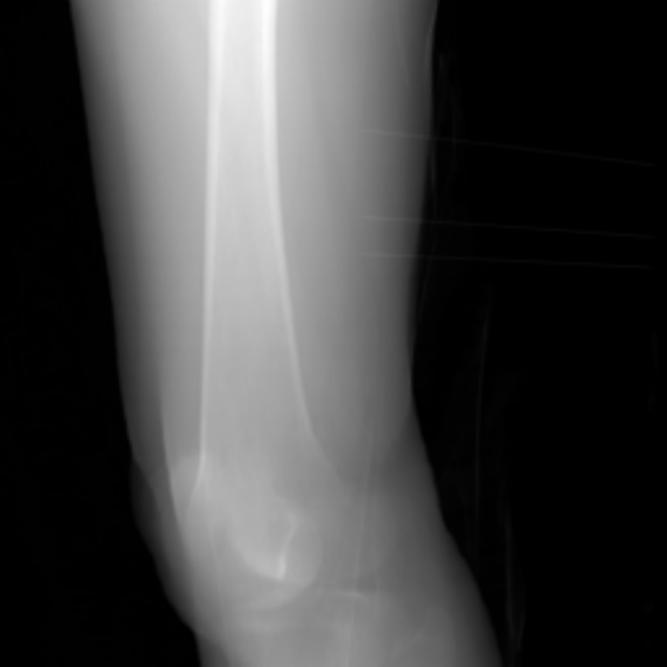}}
    \centerline{(a4)}\medskip
\end{minipage}

    \begin{minipage}[b]{0.18\linewidth}
    \centering
    \centerline{\includegraphics[width=1.6cm]{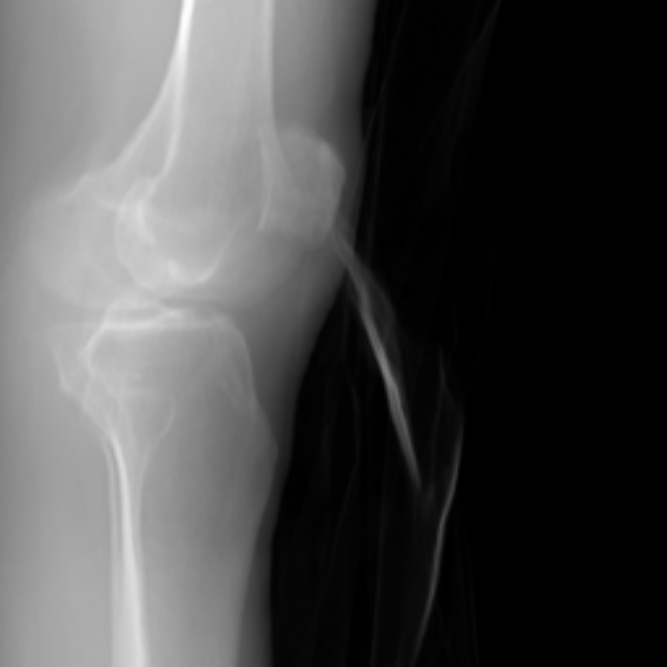}}
    \centerline{(b0)}\medskip
\end{minipage}
\hfill
    \begin{minipage}[b]{0.18\linewidth}
    \centering
    \centerline{\includegraphics[width=1.6cm]{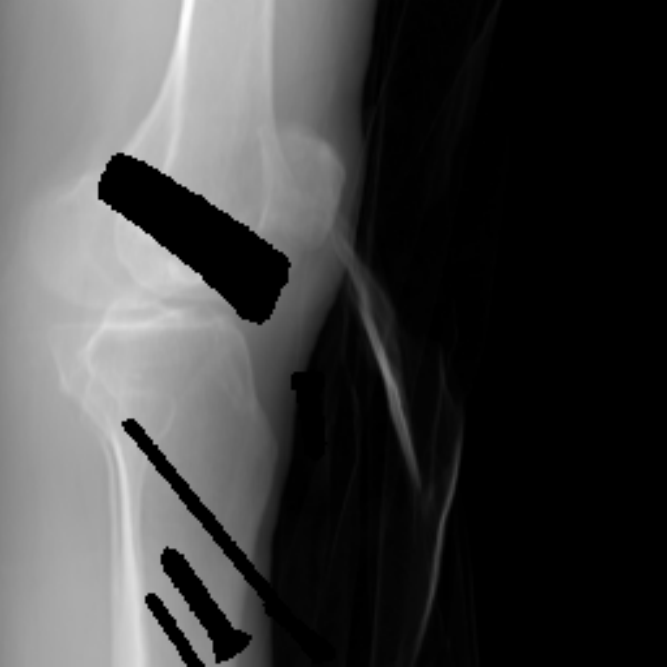}} 
    \centerline{(b1)}\medskip
\end{minipage}
\hfill
    \begin{minipage}[b]{0.18\linewidth}
    \centering
    \centerline{\includegraphics[width=1.6cm]{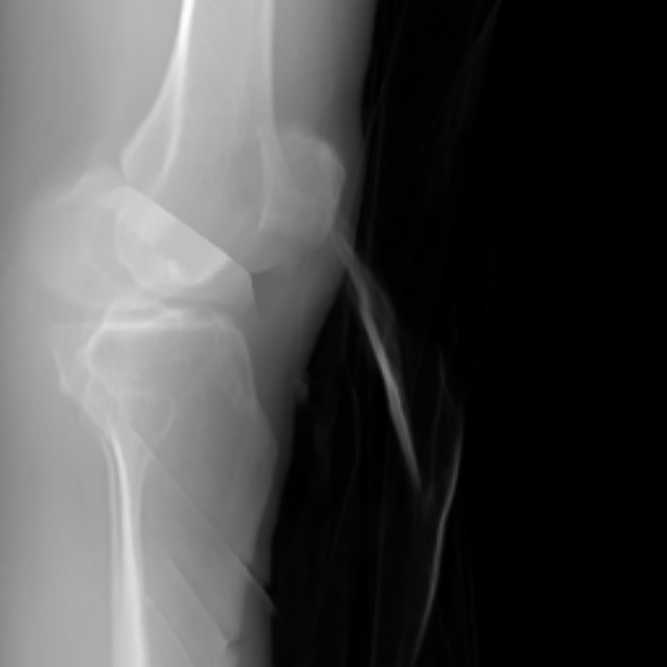}} 
    \centerline{(b2)}\medskip
\end{minipage}
\hfill
    \begin{minipage}[b]{0.18\linewidth}
    \centering
    \centerline{\includegraphics[width=1.6cm]{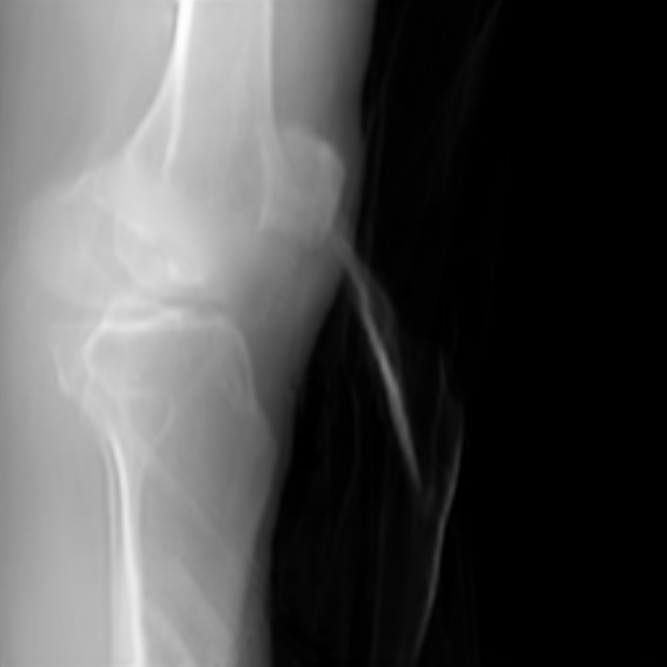}}
    \centerline{(b3)}\medskip
\end{minipage}
\hfill
    \begin{minipage}[b]{0.18\linewidth}
    \centering
    \centerline{\includegraphics[width=1.6cm]{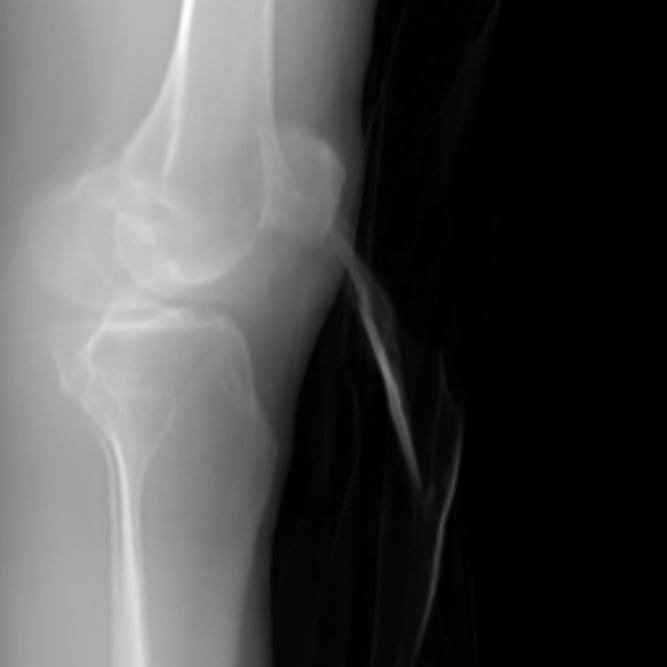}}
    \centerline{(b4)}\medskip
\end{minipage}

    \begin{minipage}[b]{0.18\linewidth}
    \centering
    \centerline{\includegraphics[width=1.6cm]{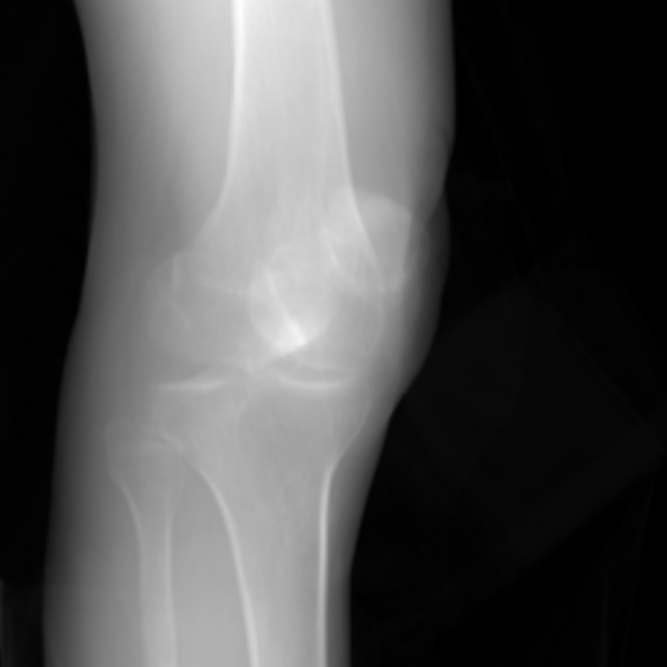}}
    \centerline{(c0)}\medskip
\end{minipage}
\hfill
    \begin{minipage}[b]{0.18\linewidth}
    \centering
    \centerline{\includegraphics[width=1.6cm]{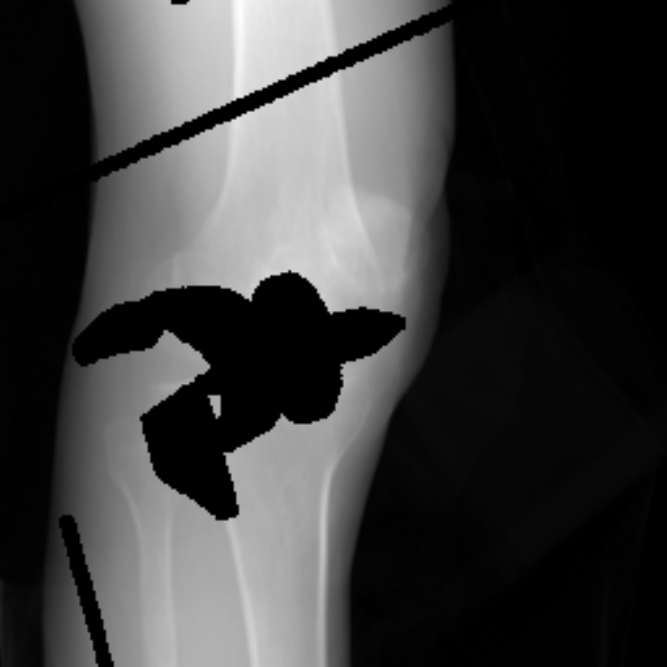}} 
    \centerline{(c1)}\medskip
\end{minipage}
\hfill
    \begin{minipage}[b]{0.18\linewidth}
    \centering
    \centerline{\includegraphics[width=1.6cm]{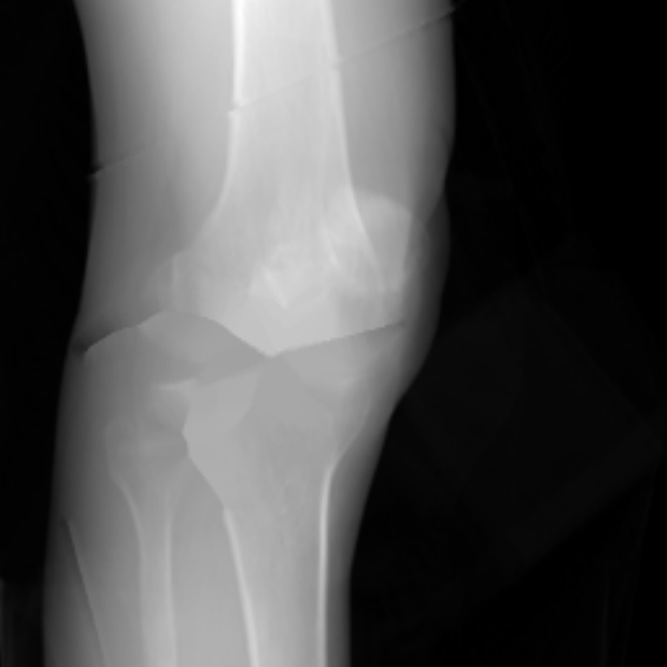}} 
    \centerline{(c2)}\medskip
\end{minipage}
\hfill
    \begin{minipage}[b]{0.18\linewidth}
    \centering
    \centerline{\includegraphics[width=1.6cm]{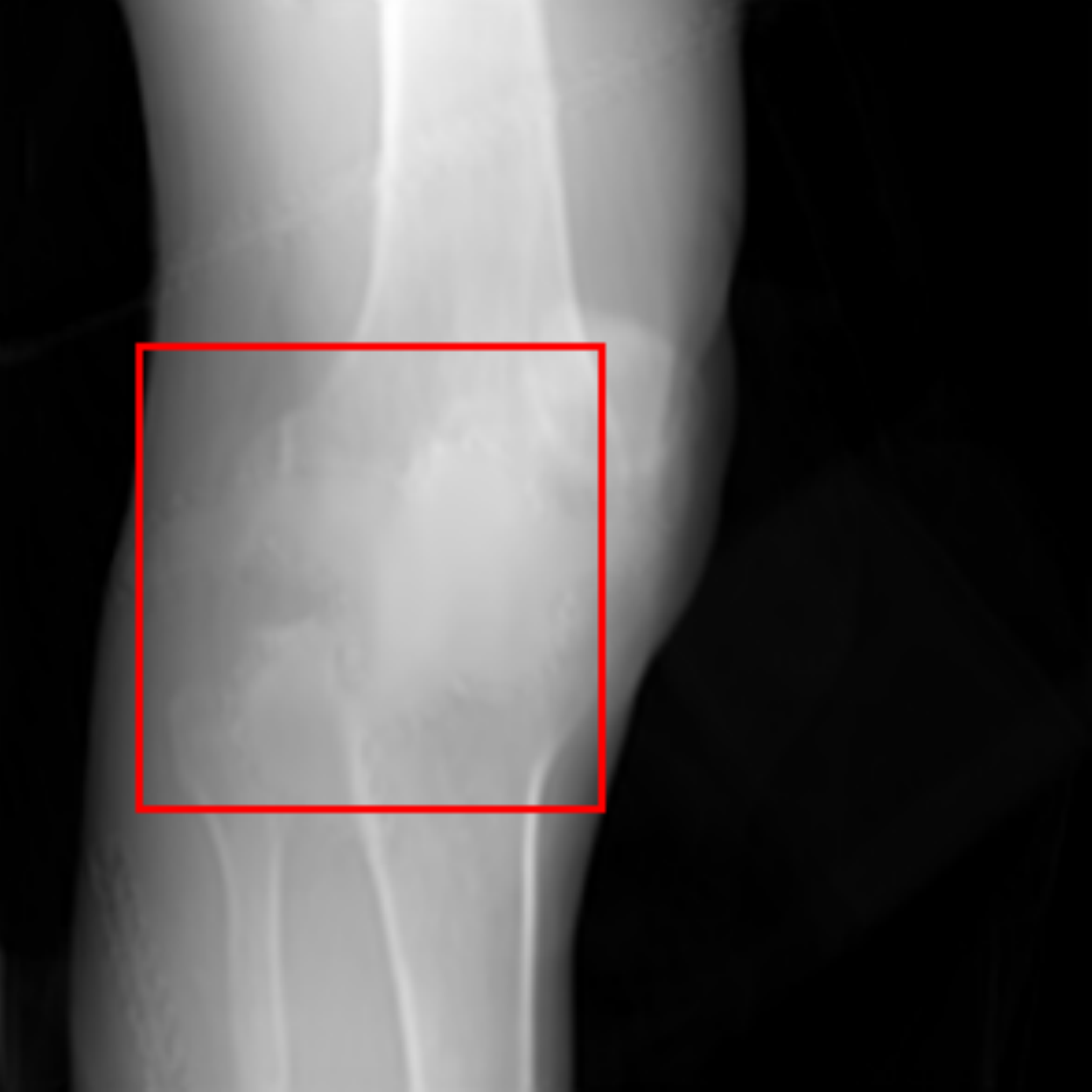}}
    \centerline{(c3)}\medskip
\end{minipage}
\hfill
    \begin{minipage}[b]{0.18\linewidth}
    \centering
    \centerline{\includegraphics[width=1.6cm]{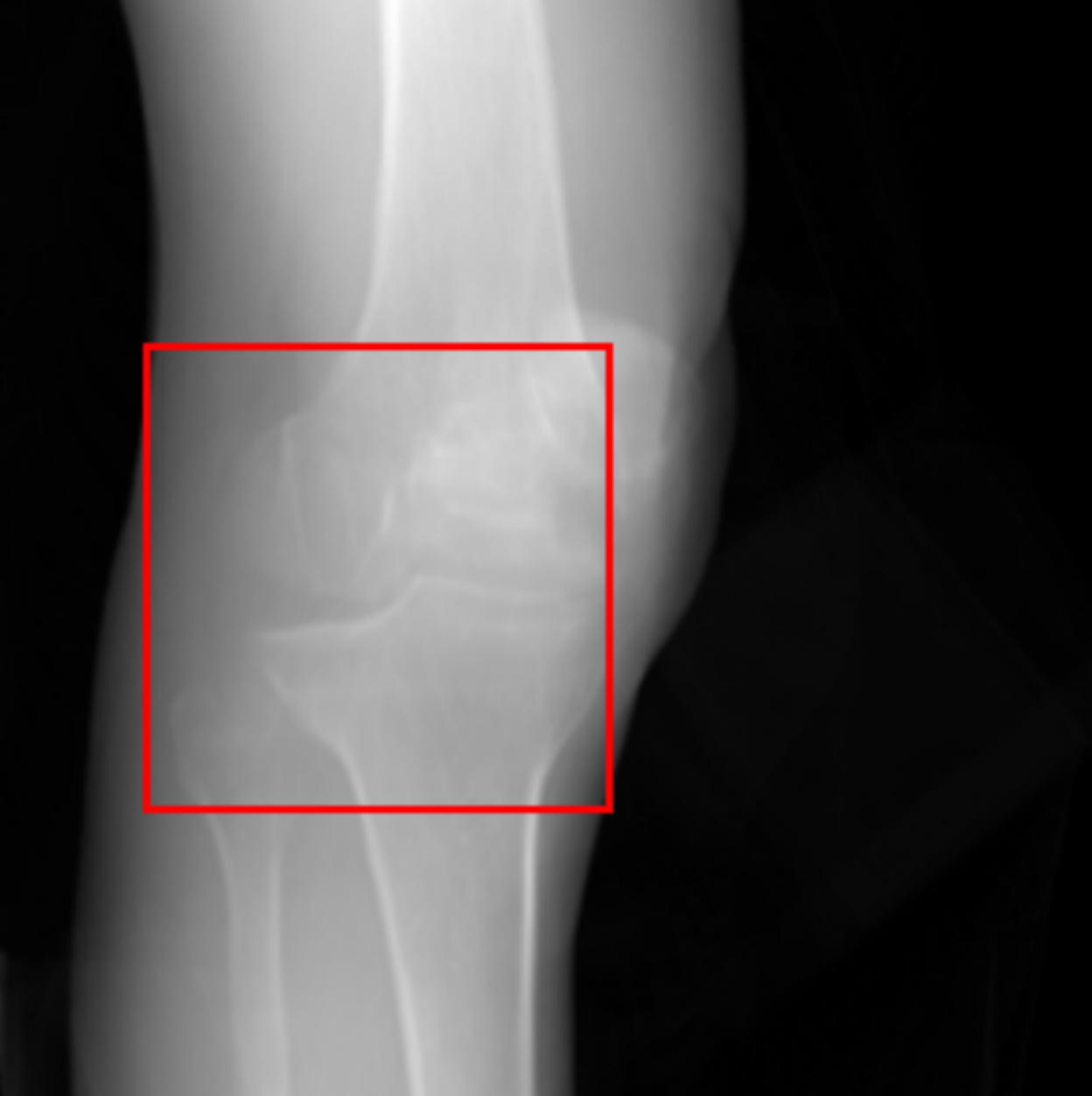}}
    \centerline{(c4)}\medskip
\end{minipage}

\caption[figure4]{\label{fig:figure4} The inpainting results under different metal masks.}
\end{figure}

In this work, result evaluation is compared among inpainted projections generated by interpolation, MPN and the score-based generative model. Three representative inpainted projections with different metal masks are shown in Fig.~\ref{fig:figure4}. From Fig.~\ref{fig:figure4}\,(a1) to (c1), more pixels are missing. The interpolation method relies on the intensity of existing pixels outside the metal area. Therefore, the inpainted projections by interpolation method have no semantic connections to the bones or soft tissue, as shown in Fig.~\ref{fig:figure4}\,(a2)-(c2). For the perspective of semantic performance, the inpainted projections by MPN can restore more details in the metal area. However, when the size of metal area increases, the prediction gets blurred, which can be observed inside the red boxes in Fig.~\ref{fig:figure4}\,(c3). In the same region in Fig.~\ref{fig:figure4}(c4), the inpainted projections by the score-based generative model have more detailed information. The quantitative evaluation results for all 300 projections with metal masks are shown in Tab.~\ref{tab:result}. The score-based generative model has the lowest mean absolute error (MAE) of 0.015 in metal regions and the highest mean peak-signal-to-noise-ratio (PSNR) of 43.00.

\begin{table*}[htb]
\caption{Quantitative results comparison.} 
\label{tab:result}
\begin{center}   
\begin{tabular}{p{3.4cm}|p{1.2cm}|p{1.2cm}|p{1.2cm}|p{1.2cm}|p{1.2cm}|p{1.2cm}} %% this creates two columns
%% use of \rule[]{}{} below opens up each row
\hline
&\multicolumn{2}{c|}{Interpolation} &
 \multicolumn{2}{c|}{MPN} &
 \multicolumn{2}{c}{Score-based model}\\
\hline
Metric & MAE &PSNR & MAE &PSNR & MAE &PSNR  \\
\hline
Metal mask & 0.031 & 35.92 & 0.025 & 39.22 & 0.015 & 43.00 \\
\hline
Circle & 0.070 & 31.50 & 0.039 & 37.89 & 0.026 & 41.31  \\
\hline
Horizontal rectangle & 0.014 & 41.11 & 0.019 & 40.10 & 0.009 & 45.51  \\
\hline
Vertical rectangle & 0.064 & 32.73 & 0.072 & 30.33 & 0.035 & 37.44 \\
\hline
\end{tabular}
\end{center}
\end{table*}

%   \begin{figure} [htb]
%   \begin{center}
%   \begin{tabular}{c} %% tabular useful for creating an array of images 
%   \includegraphics[width=7.5cm]{specialcase.pdf}
%   \end{tabular}
%   \end{center}
%   \caption[figure4] 
% %>>>> use \label inside caption to get Fig. number with \ref{}
%   { \label{fig:figure4} 
% The inpainting results under circular and rectangular masks.}
%   \end{figure}

\begin{figure}[htb]
\centering

    \begin{minipage}[t]{0.18\linewidth}
    \centering
    \centerline{Label}\medskip
\end{minipage}
\hfill
    \begin{minipage}[t]{0.18\linewidth}
    \centering
    \centerline{Input}\medskip
\end{minipage}
\hfill
    \begin{minipage}[t]{0.18\linewidth}
    \centering
    \centerline{Interpolation}\medskip
\end{minipage}
\hfill
    \begin{minipage}[t]{0.18\linewidth}
    \centering
    \centerline{MPN}\medskip
\end{minipage}
\hfill
    \begin{minipage}[t]{0.18\linewidth}
    \centering
    \centerline{Ours}\medskip
\end{minipage}

    \begin{minipage}[b]{0.18\linewidth}
    \centering
    \centerline{\includegraphics[width=1.6cm]{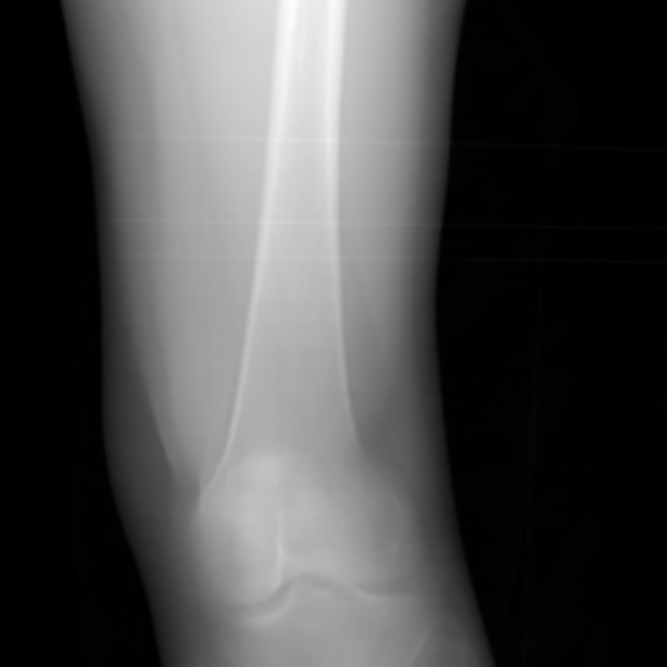}}
    \centerline{(a0)}\medskip
\end{minipage}
\hfill
    \begin{minipage}[b]{0.18\linewidth}
    \centering
    \centerline{\includegraphics[width=1.6cm]{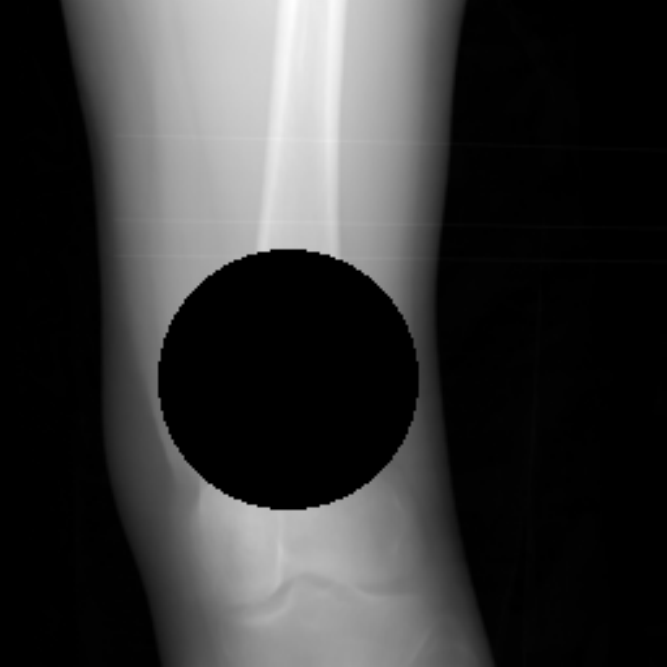}} 
    \centerline{(a1)}\medskip
\end{minipage}
\hfill
    \begin{minipage}[b]{0.18\linewidth}
    \centering
    \centerline{\includegraphics[width=1.6cm]{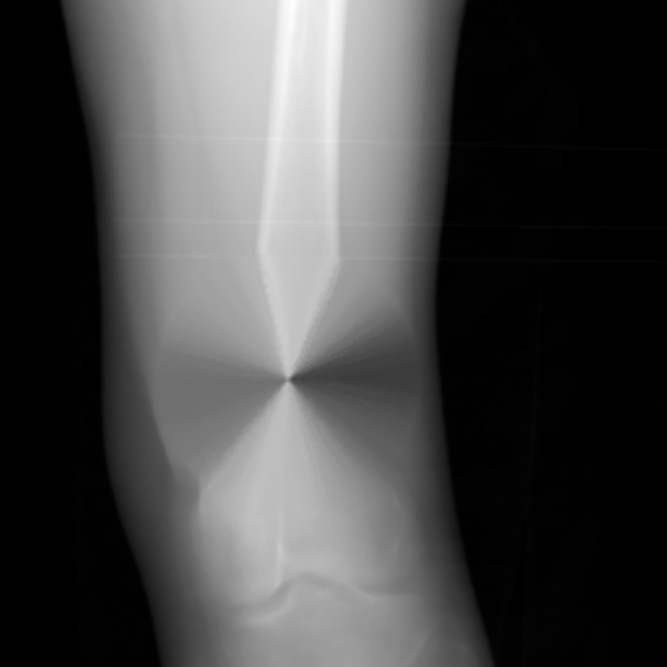}} 
    \centerline{(a2)}\medskip
\end{minipage}
\hfill
    \begin{minipage}[b]{0.18\linewidth}
    \centering
    \centerline{\includegraphics[width=1.6cm]{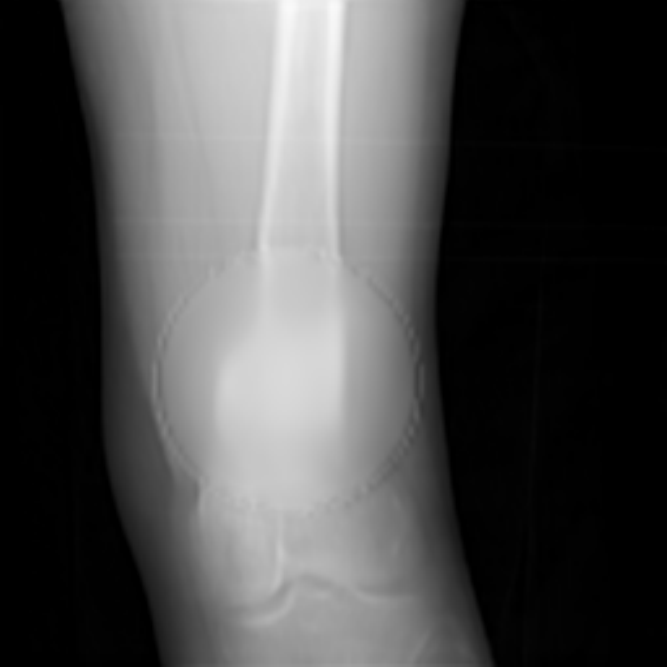}}
    \centerline{(a3)}\medskip
\end{minipage}
\hfill
    \begin{minipage}[b]{0.18\linewidth}
    \centering
    \centerline{\includegraphics[width=1.6cm]{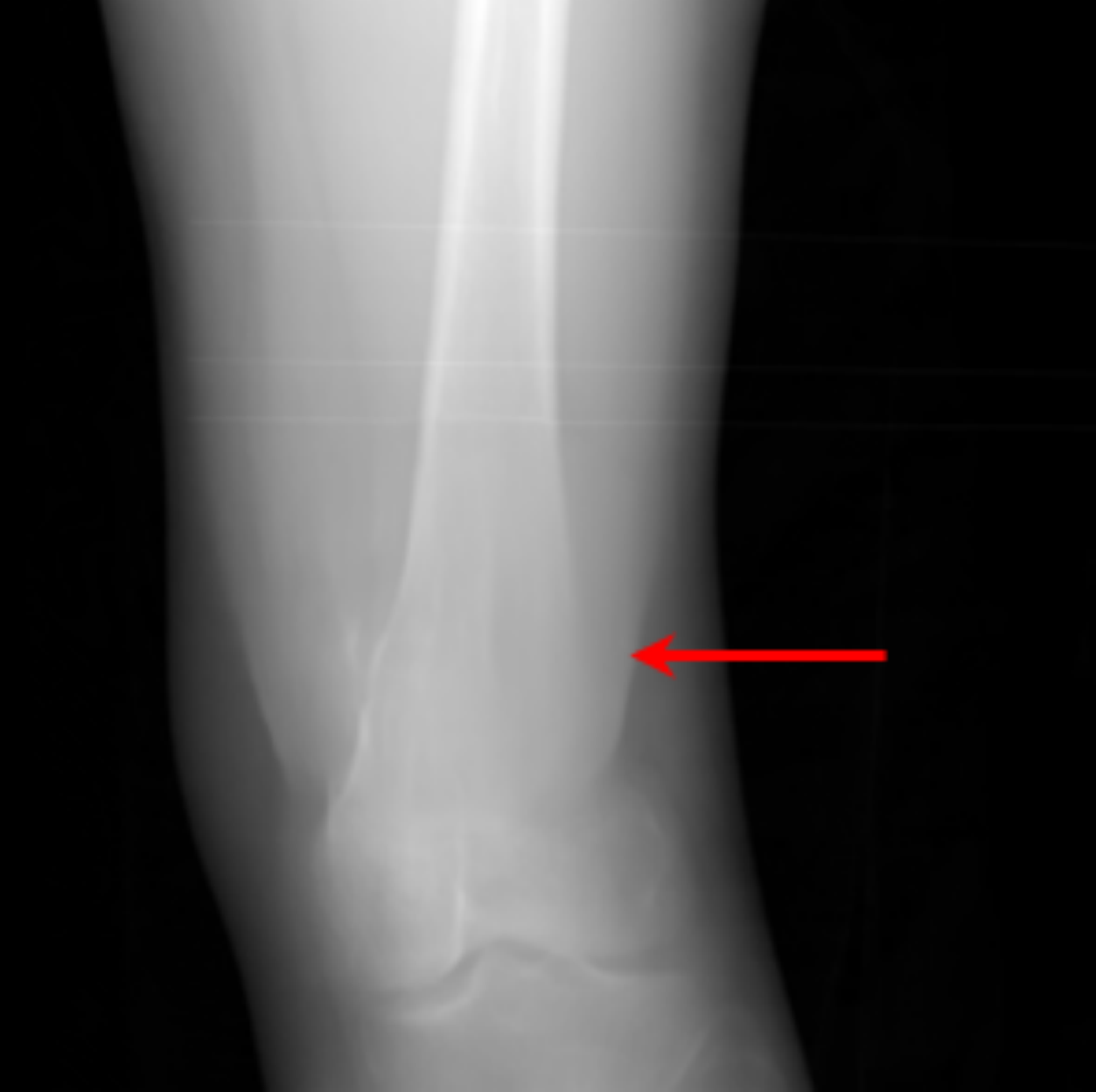}}
    \centerline{(a4)}\medskip
\end{minipage}

    \begin{minipage}[b]{0.18\linewidth}
    \centering
    \centerline{\includegraphics[width=1.6cm]{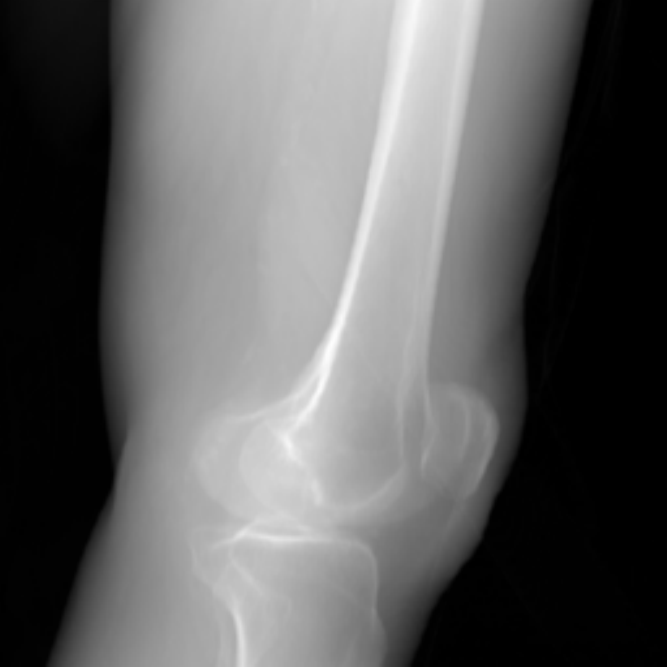}}
    \centerline{(b0)}\medskip
\end{minipage}
\hfill
    \begin{minipage}[b]{0.18\linewidth}
    \centering
    \centerline{\includegraphics[width=1.6cm]{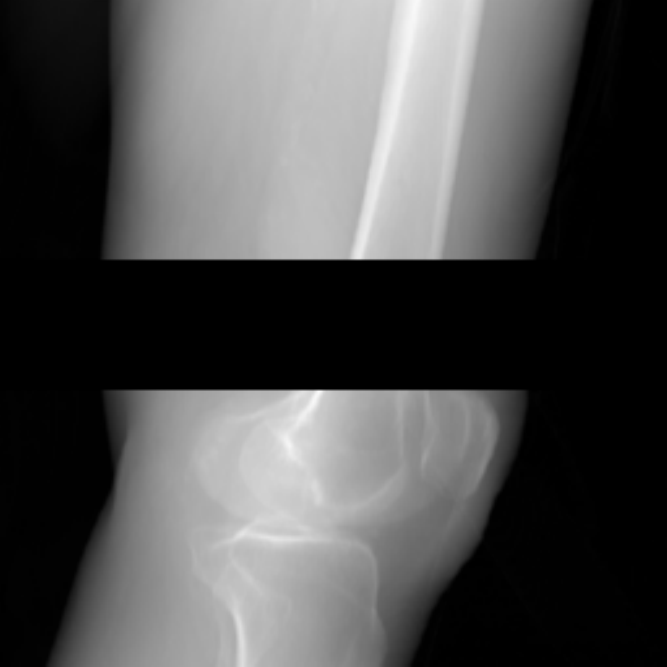}} 
    \centerline{(b1)}\medskip
\end{minipage}
\hfill
    \begin{minipage}[b]{0.18\linewidth}
    \centering
    \centerline{\includegraphics[width=1.6cm]{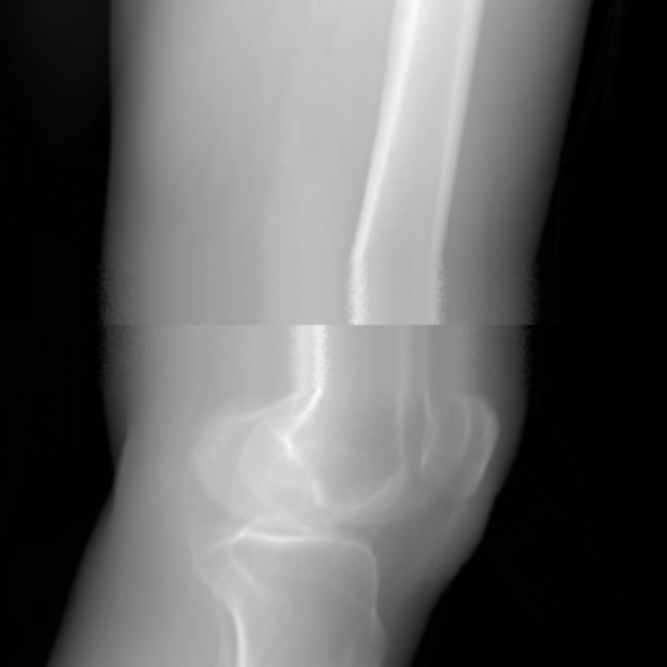}} 
    \centerline{(b2)}\medskip
\end{minipage}
\hfill
    \begin{minipage}[b]{0.18\linewidth}
    \centering
    \centerline{\includegraphics[width=1.6cm]{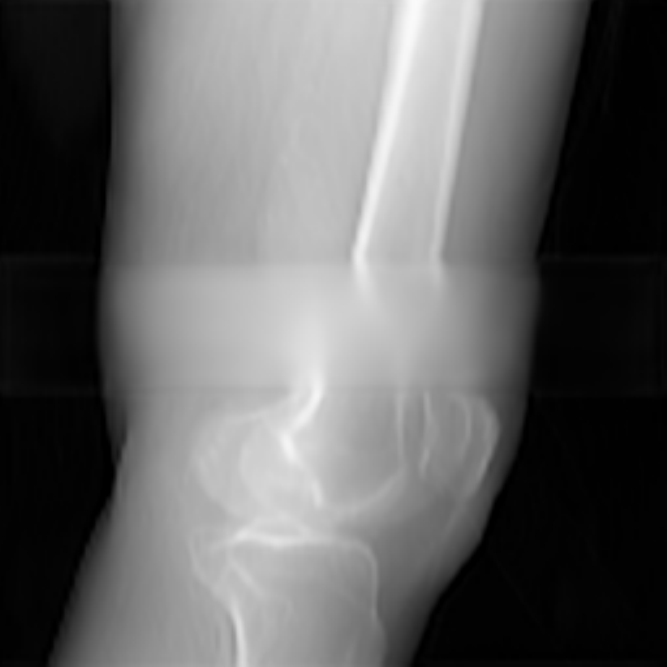}}
    \centerline{(b3)}\medskip
\end{minipage}
\hfill
    \begin{minipage}[b]{0.18\linewidth}
    \centering
    \centerline{\includegraphics[width=1.6cm]{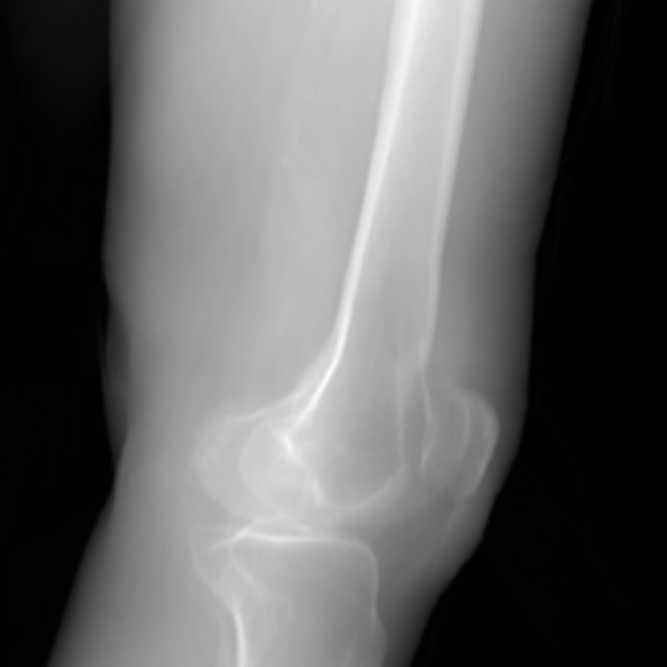}}
    \centerline{(b4)}\medskip
\end{minipage}

    \begin{minipage}[b]{0.18\linewidth}
    \centering
    \centerline{\includegraphics[width=1.6cm]{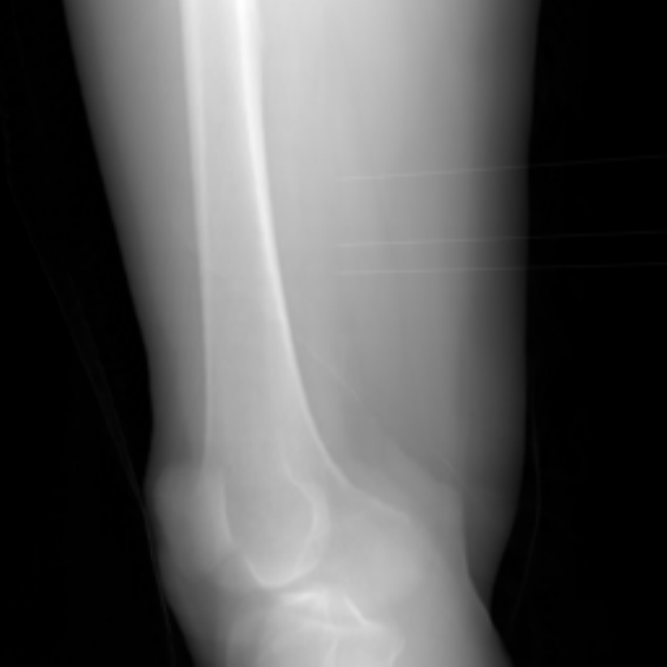}}
    \centerline{(b0)}\medskip
\end{minipage}
\hfill
    \begin{minipage}[b]{0.18\linewidth}
    \centering
    \centerline{\includegraphics[width=1.6cm]{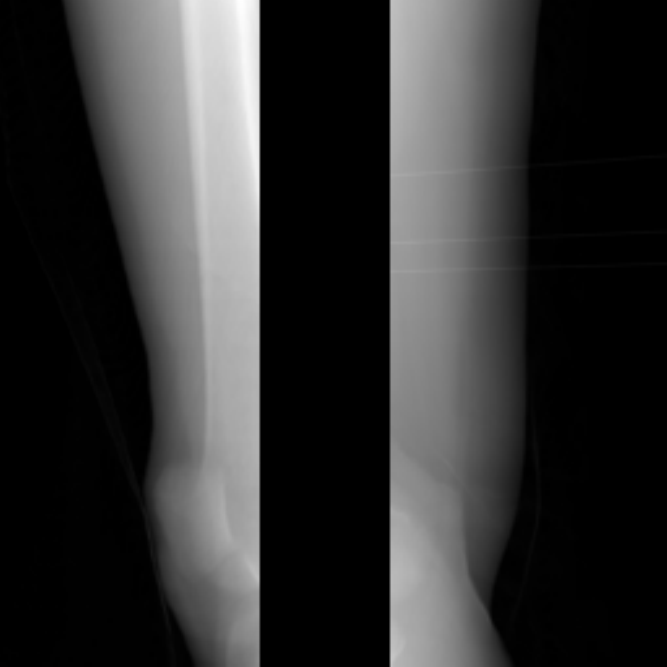}} 
    \centerline{(c1)}\medskip
\end{minipage}
\hfill
    \begin{minipage}[b]{0.18\linewidth}
    \centering
    \centerline{\includegraphics[width=1.6cm]{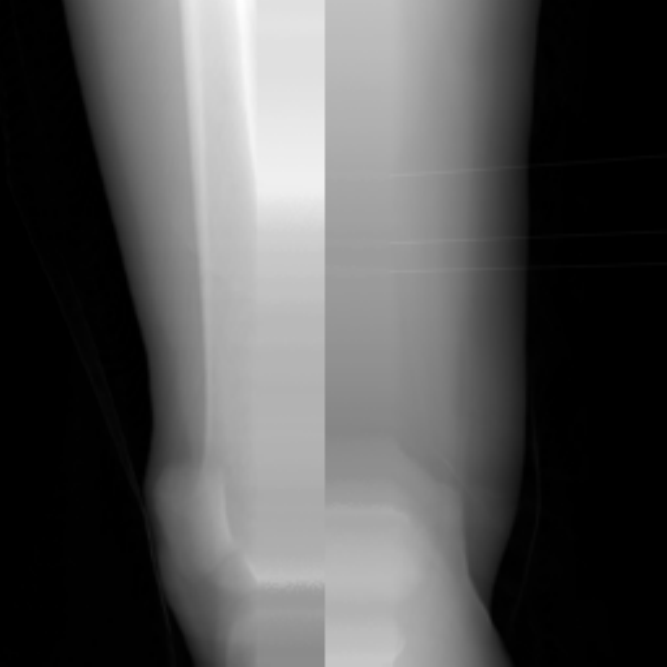}} 
    \centerline{(c2)}\medskip
\end{minipage}
\hfill
    \begin{minipage}[b]{0.18\linewidth}
    \centering
    \centerline{\includegraphics[width=1.6cm]{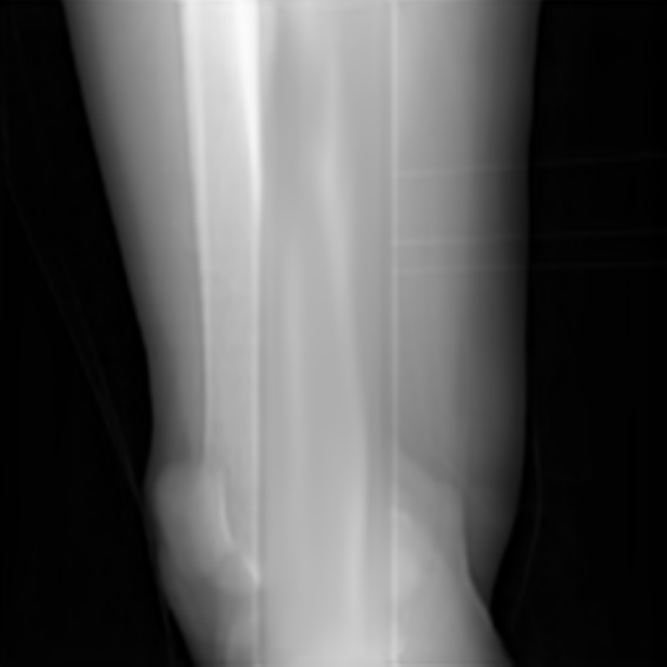}}
    \centerline{(c3)}\medskip
\end{minipage}
\hfill
    \begin{minipage}[b]{0.18\linewidth}
    \centering
    \centerline{\includegraphics[width=1.6cm]{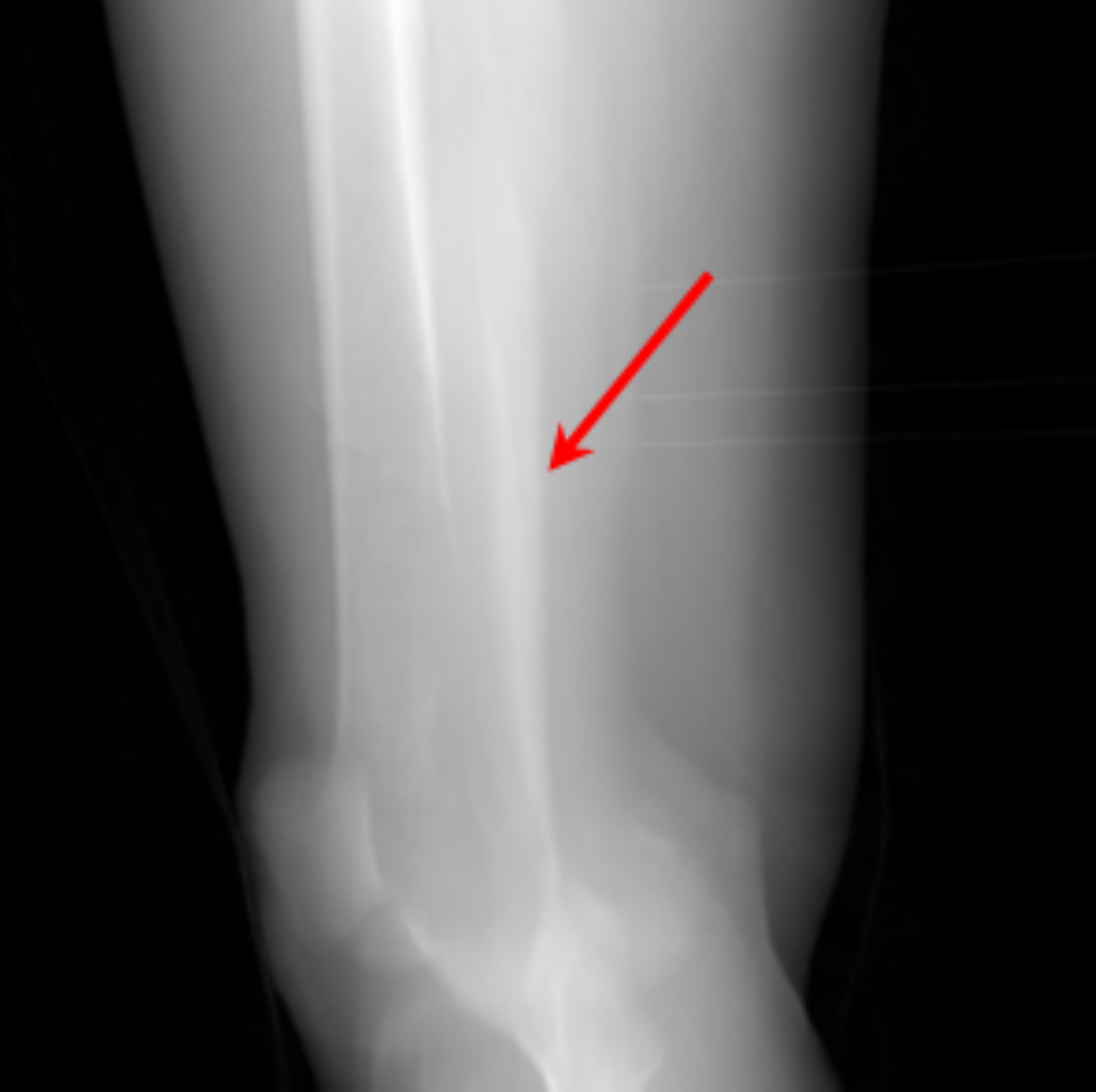}}
    \centerline{(c4)}\medskip
\end{minipage}

\caption[figure5]{\label{fig:figure5} The inpainting results under circular and rectangular masks.}
\end{figure}

Except for the metal masks, some circular and rectangular masks are also generated to test the generalization of these methods. As shown in Fig.~\ref{fig:figure5}\,(a1)-(c1), a circle, horizontal and vertical rectangles mask out the leg projections. The results by interpolation method have more discontinuity. By giving blurring predictions with distinct boundaries, MPN is not able to recover the missing pixels properly in all three cases. Although the score-based generative model also has degraded performance with blurry area in Fig.~\ref{fig:figure5}(a4) and incorrect bone in Fig.~\ref{fig:figure5}(c4), this model gives the best predictions among all methods. The quantitative results for 300 projections with circular or rectangular masks are listed in Tab.~\ref{tab:result}. The diameter of circles during test vary from 20 pixels to 60 pixels. The width of horizontal or vertical rectangle has the range of 20$-$50 pixels. In the case of rectangular masks, MPN has the largest MAE as well as the lowest PSNR. According to Tab.~\ref{tab:result}, the score-based generative model has the best results in all cases. 

MPN is trained by supervised learning, which needs the paired data of inputs and labels. This is the reason that it fails to give the correct inpainting predictions when the projections are masked out by masks which differ a lot to the training data. The score-based generative model learns the score function at different time point and it tackles the inpainting task in an unsupervised way. The background of projections guide the model to predict the missing pixels step by step in the resampling process. The metal masks in clinic have more combinations and the training data can never guarantee such diversity. Therefore, considering the robustness and generalization ability of these methods, the score-based generative model is more reliable for clinical use. However, the score-based model takes longer time because of thousands steps of resampling. The future work needs to solve this problem and projections with higher resolution should be used for clinical use.

\section{Conclusion}
This work applies the score-based generative model in metal inpainting for knee CBCT projections. By predicting more detailed information under metal masks, the proposed unsupervised method has the best performance, which is supposed to benefit MAR algorithms. What is more, the score-based generative model is able to restore the knee projections when faced with bigger circular and rectangular masks, showing its robustness in CBCT projection inpainting task.

\section{Compliance with ethical standards}
\label{sec:ethics}
This is a numerical simulation study for which no ethical approval was required.
% IEEE-ISBI supports the standard requirements on the use of animal and
% human subjects for scientific and biomedical research. For all IEEE
% ISBI papers reporting data from studies involving human and/or
% animal subjects, formal review and approval, or formal review and
% waiver, by an appropriate institutional review board or ethics
% committee is required and should be stated in the papers. For those
% investigators whose Institutions do not have formal ethics review
% committees, the principles  outlined in the Helsinki Declaration of
% 1975, as revised in 2000, should be followed.

% Reporting on compliance with ethical standards is required
% (irrespective of whether ethical approval was needed for the study) in
% the paper. Authors are responsible for correctness of the statements
% provided in the manuscript. Examples of appropriate statements
% include:
% \begin{itemize}
%   \item ``This is a numerical simulation study for which no ethical
%     approval was required.'' 
%   \item ``This research study was conducted retrospectively using
%     human subject data made available in open access by (Source
%     information). Ethical approval was not required as confirmed by
%     the license attached with the open access data.''
%     \item ``This study was performed in line with the principles of
%       the Declaration of Helsinki. Approval was granted by the Ethics
%       Committee of University B (Date.../No. ...).''
% \end{itemize}

\section{Acknowledgments}
\label{sec:acknowledgments}

The concepts and information presented in this paper are based on research and are not commercially available.

% IEEE-ISBI supports the disclosure of financial support for the project
% as well as any financial and personal relationships of the author that
% could create even the appearance of bias in the published work. The
% authors must disclose any agency or individual that provided financial
% support for the work as well as any personal or financial or
% employment relationship between any author and the sources of
% financial support for the work.

% Other types of acknowledgements can also be listed in this section.

% Reporting on real or potential conflicts of interests, or the absence
% thereof, is required in the paper. Authors are responsible for
% correctness of the statements provided in the manuscript. Examples of
% appropriate statements include:
% \begin{itemize}
%   \item ``No funding was received for conducting this study. The
%     authors have no relevant financial or non-financial interests to
%     disclose.'' 
%   \item ``This work was supported by […] (Grant numbers) and
%     […]. Author X has served on advisory boards for Company Y.'' 
%   \item ``Author X is partially funded by Y. Author Z is a Founder and
%     Director for Company C.''
% \end{itemize}

\bibliographystyle{IEEEbib}
\bibliography{strings,refs}

\end{document}